\documentclass[aip,cha]{revtex4-1}
\usepackage{amsmath,amssymb}
\usepackage{epsfig}
\usepackage{graphicx}
\usepackage{dcolumn}
\usepackage{bm}
\usepackage{times}
\usepackage{xcolor}

\begin{document}


\title{Network Inoculation: Heteroclinics and phase transitions in an epidemic model}

\author{Hui Yang}\email{jenifferyang25@gmail.com}
\affiliation{Web Sciences Center, University of Electronic
Science and Technology of China, Chengdu 611731, China}
\affiliation{Department of Engineering Mathematics, University of Bristol, Bristol BS8 1UB, UK}

\author{Tim Rogers}
\affiliation{Centre for Networks and Collective Behaviour, Department of Mathematical Sciences, University of Bath, Claverton Down, BA2 7AY, Bath, UK}
\author{Thilo Gross}
\affiliation{Department of Engineering Mathematics, University of Bristol, Bristol BS8 1UB, UK}

\date{\today}

\begin{abstract}
In epidemiological modelling, dynamics on networks, and in particular adaptive and heterogeneous networks have recently received much interest. Here we present a detailed analysis of a previously proposed model that combines heterogeneity in the individuals 
with adaptive rewiring of the network structure in response to a disease. We show that in this 
model qualitative changes in the dynamics occur in two phase transitions. In a macroscopic 
description one of these corresponds to a local bifurcation whereas the other one corresponds 
to a non-local heteroclinic bifurcation. This model thus provides a rare example of a system 
where a phase transition is caused by a non-local bifurcation, while both micro- and macro-level dynamics are accessible to mathematical analysis. The bifurcation points mark the onset of a behaviour that we call \emph{network inoculation}. In the respective parameter region exposure of the system to a pathogen will lead to an outbreak that collapses, but leaves the network in a configuration where the disease cannot reinvade, despite every agent returning to the susceptible class. We argue that this behaviour and the associated phase transitions can be expected to occur in a wide class of models of sufficient complexity.
\end{abstract}

\maketitle

\textbf{Throughout history epidemic diseases have been a major cause of death in the human population. After a brief respite during the mid twentieth centrury, incidences of epidemics are now on the rise again, due to the emergence of new diseases such as 
Aids and Ebola, and the return of old killers, such as Tuberculosis and Influenza. 
Consequently, the study of epidemilogy has received much recent attention from the mathematics 
and physics communities. In particular, network models provide a new theoretical tool by which the 
spreading of epidemic diseases can be understood and lessons for the real world can be learned. The present direction of this field is to push network models to greater realism by 
incorporating more and more aspects of real world epidemics, while maintaining mathematical 
and/or numerical tractability of the models. In this paper we study the combined effect of
two properties of real world contact networks across which real epidemics spread: adaptivity and heterogeneity. The network is adaptive in the sense that individuals in the network can respond to the presence of the disease, and it is heterogeneous in the sense that the individuals 
represented by network nodes have different properties, making them more or less susceptible to the disease. We show that combining these features leads to a phenomenon that we call \emph{network inoculation}. Exposure of a given initial network to a pathogen can lead to an outbreak that collapses and leaves the network resistant to future outbreaks. This 
resistance is acquired solely through the rewiring of network structure, without any becoming physically immune to the disease. We use a variety of tools, including agent-based simulation, moment expansions, percolation methods, and numerical continuation, to reveal the heteroclinic mechanism that leads to this inoculation phenomenon.       
}

\section{Introduction}
A central goal in complex systems research is to understand how macroscopic transitions 
arise from the microscopic interactions within a system~\cite{sethna2006statistical}. In this context an important role is played by coarse-grained models, describing the system in terms of a set of ordinary differential equations (ODEs)~\cite{kevrekidis2004equation,gross2008robust}. By capturing the dynamics of the system in terms of a suitable set of  variables, it is sometimes possible to construct a faithful model of a given transition that is easy enough to be tractable by the tools of nonlinear dynamics. In the analysis the transition then appears as a bifurcation, whose study reveals deep insights into the nature and behaviour of the underlying microscopic system.    

A paradigmatic example is the epidemiological SIS model~\cite{anderson1991infectious}. In its simplest incarnation, this model describes the propagation of an infectious disease in a group of randomly interacting agents. Each agent is either infected with the disease (state I) or susceptible to the disease (state S). In time, the state of agents changes due to transmission of the disease and recovery of infected agents. The dynamics of this system can be understood by writing a single differential equation that captures the proportion of agents $[I]$ that are infected. Depending on the details of interactions the system either approaches a state where the disease is extinct or a state where it persists at a constant level. In the ODE-based model, the transition between the two qualitatively different types of behaviours occurs at a threshold parameter value that is a bifurcation point.  

In the SIS model, and many other models besides, the important bifurcation is local, i.e.~it is a bifurcation that can be characterised by changes in the phase portrait in the proximity of a single steady state or other invariant set~\cite{kuznetsov2013elements}. For instance in the epidemic example this bifurcation is a transcritical bifurcation in which a steady state with non-zero density of infected agents intersects the state where the disease is extinct, and the two exchange their stability. Thus the relevant changes in the phase portrait occur in the vicinity of the extinct steady state.   

The transcritical bifurcation and its close relatives, the fold and pitchfork bifurcations have been linked to phase transitions in a wide variety of systems including
epidemics~\cite{anderson1991infectious}, collective motion of animals~\cite{couzin2011uninformed,huepe2011adaptive}, human opinion formation~\cite{tanabe2013complex,bagnoli2013topological}, neuronal dynamics~\cite{shayer2000stability,huber2005cooperative} and others. In a smaller number of models the underlying bifurcation is a Hopf bifurcation, which marks the onset of, at least transient, oscillations~\cite{risler2004universal,gross2006epidemic,acebron2005kuramoto}. However, even the Hopf bifurcation is a local bifurcation. By comparison models in which 
a phase-transition corresponds to a non-local bifurcation in a macroscopic model are rare. 

\begin{figure}
\begin{center}
\epsfig{file=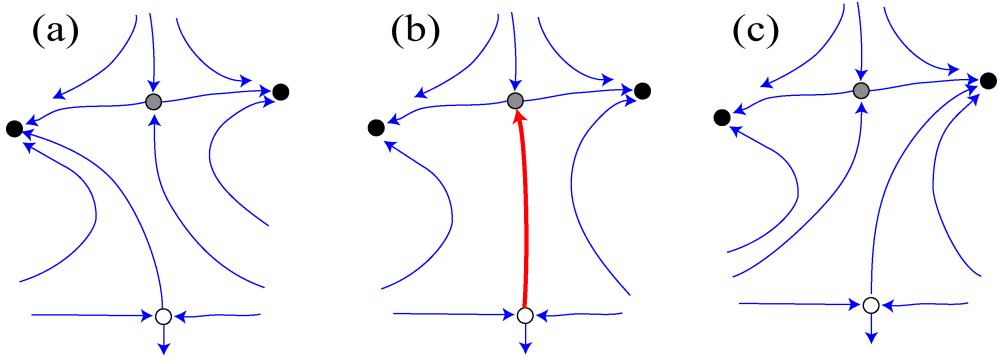,width=0.8\linewidth}
\caption{(Color Online) Sketch of the phase portrait before, during, and after a heteroclinic bifurcation. In the system two attractors (black dots) coexist with two saddles (white and grey dots). The flow field is indicated by thin blue arrows. Before the bifurcation, a small perturbation launch the system on a trajectory leading to the left attractor (a). As parameters are changed a heteroclinic connection between the saddles is formed, shown by red strong arrow in (b). After the bifurcation fluctuations on the white saddle can now lead to a final state at the right attractor (c), while the left attractor has become unreachable from the white saddle.}\label{Figure1}
\end{center}
\end{figure}

In nonlinear dynamics several non-local bifurcations have been described. An example of particular interest for the present paper is the heteroclinic bifurcation~\cite{guckenheimer1983nonlinear,shilnikov2001methods,ashwin2005nonlinear}. In this bifurcation a transition in the macroscopic dynamics of a system occurs, due to the appearance of a trajectory connecting different invariant sets (see Fig.~\ref{Figure1}). Such bifurcations already occur robustly in relatively low-dimensional dynamical systems~\cite{ashwin2005nonlinear}. The closely related homoclinic bifurcation often marks the point where 
a limit cycle is destroyed and thus causes a discontinuous phase transitions in many systems. One of these is the adaptive SIS model: an SIS system, where additionally the susceptible nodes try to avoid infection by rewiring their links away from infected nodes~\cite{gross2006epidemic}. 

In the adaptive SIS model the importance of the homoclinic bifurcation is very minor. The bifurcation can play the role of an epidemic threshold 
in a small parameter space, where it occurs close to a Hopf bifurcation, such that very large simulations are needed to see the limit cycle. A homoclinic bifurcation was also found to enable a transition to full cooperation in a game theoretical model, but required global information transfer between agents~\cite{zschaler2010homoclinic}.   

Homoclinics, heteroclinics, and other non-local bifurcations are also known to play a major role in fluid dynamics and climate system modelling~\cite{pikovsky2003synchronization,titz2002homoclinic}. Perhaps the best known example is the Lorenz model~\cite{lorenz1963deterministic}. However, this model is directly formulated on the macroscopic level, such that no direct connection to the phase transition in the underlying microscopic dynamics can be made. By contrast, models that resolve the detailed dynamics are often too complex to reveal a detailed picture of heteroclinics in the dynamics by use of bifurcation theory. 

In a recent paper we investigated the dynamics of a heterogeneous adaptive SIS model, 
which combined SIS dynamics and disease avoidance behaviour with heterogeneity in the 
susceptibility of the population. Both heterogeneity and adaptivity are known to impact 
the dynamics of diseases of humans~\cite{dye2003modeling}, and are therefore presently high on the agenda in network epidemiology. For instance adaptivity was shown to significantly increase the epidemic threshold and lead to a first-order transition at the onset of the disease~\cite{shaw2008fluctuating, zanette2008infection, marceau2010adaptive, wang2011epidemic,shaw2012epidemic} and can induce robust oscillations~\cite{gross2008robust}. 
Moreover, studies showed that adaptive disease avoidance behaviour can effectively enhance the impact of disease control efforts~\cite{risau2009contact,shaw2010enhanced,van2010adaptive,yang2012efficient}.
The heterogeneity between individuals was shown to lower the epidemic threshold in some networks~\cite{pastor2001epidemic,parshani2010epidemic,smilkov2014beyond}, but can also reduce the size and risk of outbreaks~\cite{miller2007epidemic,miller2008bounding,neri2011heterogeneity,neri2011effect,katriel2012size,smilkov2014beyond}

In~\cite{yang2015large} we found that a plausible disease avoidance mechanism can lead 
to states where the network has a heterogeneous topology, but is more resilient to the 
invasion of diseases than it would be possible in less heterogeneous topologies. 
These findings are thus contrary to the intuition gained from landmark results for simpler models~\cite{albert2000error,pastor2001epidemic}, which seem to suggest that 
heterogeneous topologies would always aid the transmission of the disease.

While our previous publication~\cite{yang2015large} pointed to a mechanism that leads to the emergence of extraordinarily stable heterogeneous topologies, the actual transition at which this mechanisms sets in was too complicated to analyze within the scope of that paper. 
Here we investigate this transition first in the previously proposed model and then in a highly stylized model that enables a deeper understanding of the phenomenon. 

We find that the threshold for the onset of an endemic infection does not correspond to a loss of 
stability of the disease-free state. Instead, there is a large parameter range in which initial disease-free networks are unstable and thus permit disease invasion, but outbreaks do not lead to an endemic state but collapse back to another disease-free state, with different network topology. The dynamics of the system in this region is thus reminiscent of an 
SIR model. However, there is no recovered (R) agent state in the model that confers immunity. 
Instead, an initial outbreak leads to the formation of more resilient network topologies, and thus ``inocculates'' the network against future disease invasion. 

Network inocculation is characterized by the presence of heteroclinic orbits that connect different disease free states. Because of the basic physics of the system the disease-free states form a manifold. When the infectivity of the disease is changed the orbit starting from a given initial steady state may connect to a (unique) saddle point. When this happens a saddle-heteroclinic bifurcation occurs, which ends the inocculation-type dynamics from the respective initial network. For all higher values of infectivitity the heteroclinic trajectory from that initial state leads to an endemic state where the disease can persist in the system indefinitely. Thus the onset of endemic disease dynamics is marked by a phase transition caused by a heteroclinic bifurcation in the underlying dynamics.   

This paper is organized as follows: We start by reviewing the previously proposed model (Sec.~\ref{secModel}). In agent-based simulations we observe that the outcomes of simulation runs can be classified into 3 different types (Sec.~\ref{secClassification}). We then explore the phase boundaries between the three different types of outcomes. Using percolation theory we analytically compute the threshold where outbreaks start to occur (Sec.~\ref{secPercolation}). Thereafter, using moment expansions, we formulate a macroscopic  model of the dynamics in terms of ordinary differential equations (Sec.~\ref{secMoments}), this model allows us to study the dynamics by tools of dynamical systems theory. Combining, results from all of the tools established up to this point we show that the transition from outbreaks to endemic behavior occurs due to a heteroclinic bifurcation (Sec.~\ref{secTransition}). To understand this transition in greater detail we finish by formulating and analzing a simpler solvable model for the network inocculation phenomenon (Sec.~\ref{secSolvable}).      

\section{Heterogeneous Adaptive SIS Model \label{secModel}}
We consider a population of $N$ agents, which can be either infected (state I) or susceptible to the disease (state S).
The agents are connected by a total of $K$ bilateral social contacts. Thus the system can be described as a network in which the the agents are the network nodes and 
the social contacts are the links. In time the system evolves (a) because of the epidemic dynamics, and (b) due to a behavioural response of the agents to the disease, which leads to the rewiring of links. 

During the course of the epidemic dynamics (a) for every link connecting a susceptible and an infected agent there is a chance that the susceptible agent becomes infected, amounting to an infection rate of $\beta\psi$ (per link),
where $\beta$ is a parameter that controls the overall infectivity of the disease and $\psi$ 
is a parameter that describes the susceptibility of the susceptible agent. In particular, we consider the case where two types of agents exist: highly susceptible agents (type A) and less susceptible agents (type B). These types are intrinsic properties of the agents, i.e.~unlike the epidemic states the type of an agent never changes. Furthermore, all infected agents recover at a fixed rate $\mu$, which is identical for all agents. Upon recovery, agents immediately become susceptible again. 

We denote the proportion of agents of type A in the population by $p_{\rm a}$ and their susceptibility by $\psi_{\rm a}$. The remaining portion of agents $p_{\rm b}=1-p_{\rm a}$ is of type B and has susceptibility $\psi_{\rm b}<\psi_{\rm a}$. In the following we chose these parameters such that  $p_{\rm a}\psi_{\rm a} + p_{\rm b} \psi_{\rm b} = \langle \psi \rangle=0.5$. We thus control the heterogeneity of susceptibility in the population by changing $\psi_{\rm a}$ and $\psi_{\rm b}$ simultaneously such that the mean 
susceptibility $\langle \psi \rangle$ remains fixed. Hence the intra-individual heterogeneity is indicated by one of the parameters, say $\psi_{\rm a}$, whereas the overall spreading rate is controlled by the epidemic parameter $\beta$.

In the social dynamics (b), the agents react to the presence of the disease by rewiring their social connections. 
In each small time interval of length $dt$, a susceptible agent who is linked to an infected agent breaks that link with probability $\omega dt$. For every link a susceptible agent breaks it and establishes a new link to a randomly chosen susceptible agent, such that the total number of links is conserved. 
 
In the following, we use the parameters $ N=10^{5}$, $ K=10^{6}$, $\omega=0.2$, $\mu=0.002$ and $\langle \psi \rangle=0.5$ unless noted otherwise.

\section{Classification of Outcomes \label{secClassification}}
We start the analysis by numerically exploring the possible outcomes in agent-based simulations. We initialize the system as a Erd\H{o}s-R\'{e}nyi random graph, where 
each agent is initially infected with probability $i_0=0.0002$ and susceptible otherwise.
We then simulate the time evolution of the system of agents using a Gillespie algorithm~\cite{zschaler2013largenet2}. 

\begin{figure}
\begin{center}
\epsfig{file=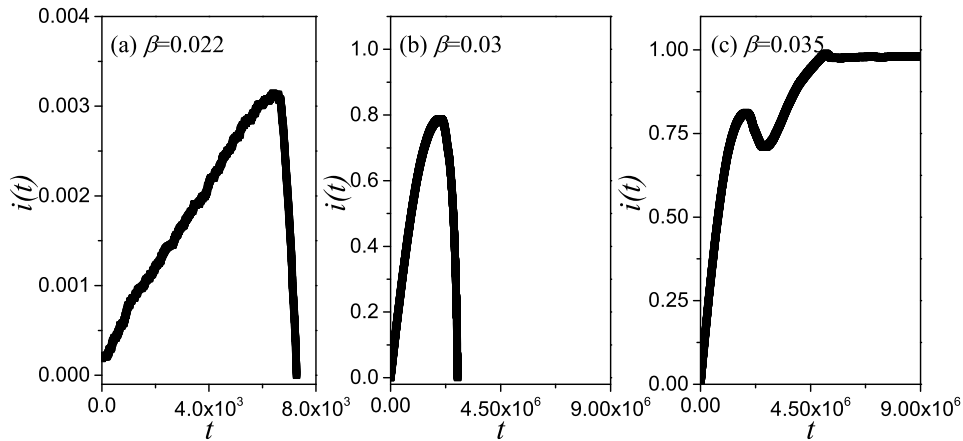,width=0.8\linewidth}
\caption{(Color online) Three typical timeseries from agent-based simulations. If the infectivity is low (left, $\beta=0.022$) then the epidemic dies out quickly and the system freezes in the disease-free state (note the different axis scaling on this plot). For intermediate infectivity (center, $\beta=0.03$) there is an initial outbreak, which infects a large proportion of the agents. 
However, subsequently this outbreak collapses and the system once again approaches the disease 
free state. If infectivity is high (right, $\beta=0.035$) then the system approaches an endemic state where the diseases remains in the system in the long term. Parameters: $\psi_a=0.65$, $\psi_b=0.05$, $w=0.2$, $ \mu=0.002 $, $i_0=0.0002$, $N = 10^5$, $K = 10^6$. }\label{Figure2}
\end{center}
\end{figure}

Three typical outcomes are shown in Fig.~\ref{Figure2}. Depending on the parameter values, we observe either a rapid collapse to a disease-free state, before a significant proportion of the agents have been infected (type I), an initial epidemic outbreak, in which a large proportion of agents are infected (type II), or an outbreak leading to an endemic state where the disease persists indefinitely (type III).  

Let us try to extrapolate from the finite-size simulation to arbitrarily 
large systems. The results of this analysis should hold in large finite systems encountered in the real world or studied in large agent-based simulations, where finite size effects are mostly irrelevant, due to the size of the system considered.

Referring to an infinitely large system is attractive because it allows us to avoid problems in the classification of behaviours that exist in the finite system. Consider that in the finite case the difference between type I (recovery to the disease-free state) behaviour  and type II (outbreak, collapse) behaviour is not rigorously defined,  i.e.~the transition is gradual as the number of infected at maximum increases. Furthermore, even the difference between type II and type III (persistent) behaviour becomes fuzzy: The finite size agent-based simulation has a finite probability to spontaneously collapse to the absorbing disease-free state. Thus persistent dynamics cannot be a true long-term behaviour, although we never observed such a collapse of apparently persistent epidemics in all but the smallest simulation runs (e.g.~$N<100$) or when the system is just at the epidemic threshold.  

\begin{figure}
\begin{center}
\epsfig{file=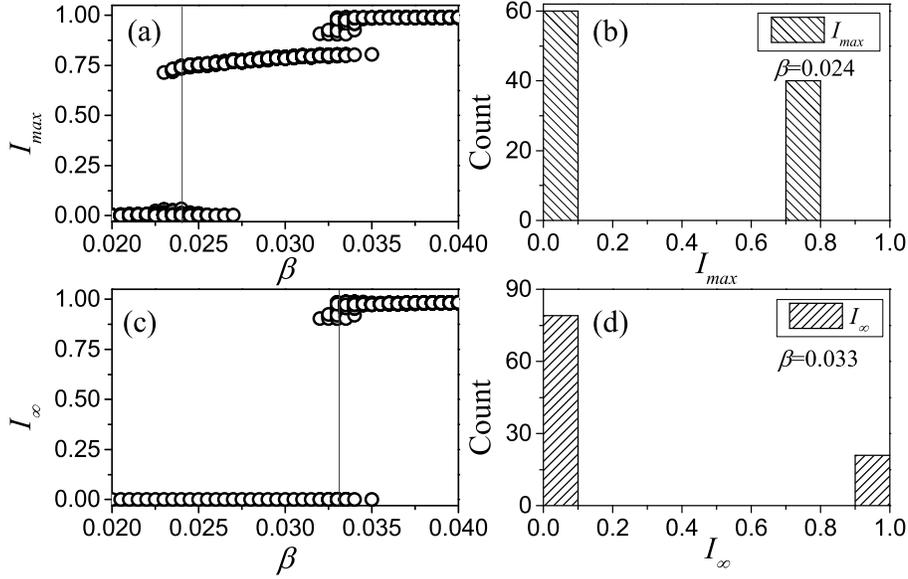,width=0.8\linewidth}
\caption{ Classification of outcomes from agent-based simulations. Shown are the maximal proportion of infected agents encountered in a simulation run, $I_{\rm max}$ (top left) and the proportion of infected after long time, $I_{\infty}$ ($t=10^7$, bottom left). The symbols represent observed outcomes for each of 100 simulation runs for each value of infectivity $\beta$, many of which are so similar that they are indistinguishable. 
It is apparent that three qualitatively different outcomes are observed: $I_{\infty}\approx 0$, $I_{\rm max}\approx 0$ (type I), $I_{\infty}\approx 0$, $I_{\rm max}> 0$ (type II), $I_{\infty}> 0$, $I_{\rm max}> 0$ (type III). While two different outcomes are possible for some values of $\beta$, they can be clearly distinguished in this case, see Histograms in the panels on the right, with values of $\beta$ corresponding to the thin lines shown in the left plots.   
Parameters: $\psi_a=0.65$, $\psi_b=0.05$, $w=0.2$, $ \mu=0.002 $, $i_0=0.0002$, $N = 10^5$, $K = 10^6$ }\label{Figure3}
\end{center}
\end{figure}
  
By contrast, the different types of behaviour can be cleanly defined in the infinite system. 
We say that the behavior of the system is of type I, if the epidemic never grows to a point where a finite proportion of the agents is infected. This makes type I behavior qualitatively 
different from type II and type III, where the at some point a finite proportion of the agents is infected. We further distinguish type II and type III behaviour by their long-term behavior: We can say that a system shows type III behavior if in the infinite size limit, a finite proportion of the agents are infected after arbitrarily long time.   

Now returning to finite systems, the considerations above enable us to classify the dynamics using scaling relationships. However, in practice this is not necessary as the differences in sufficiently large simulations are clear cut. Results from simulations with $N=10^5$ nodes in Fig.~\ref{Figure3} show that the three types of outcomes can be clearly distinguished.

\begin{figure}
\begin{center}
\epsfig{file=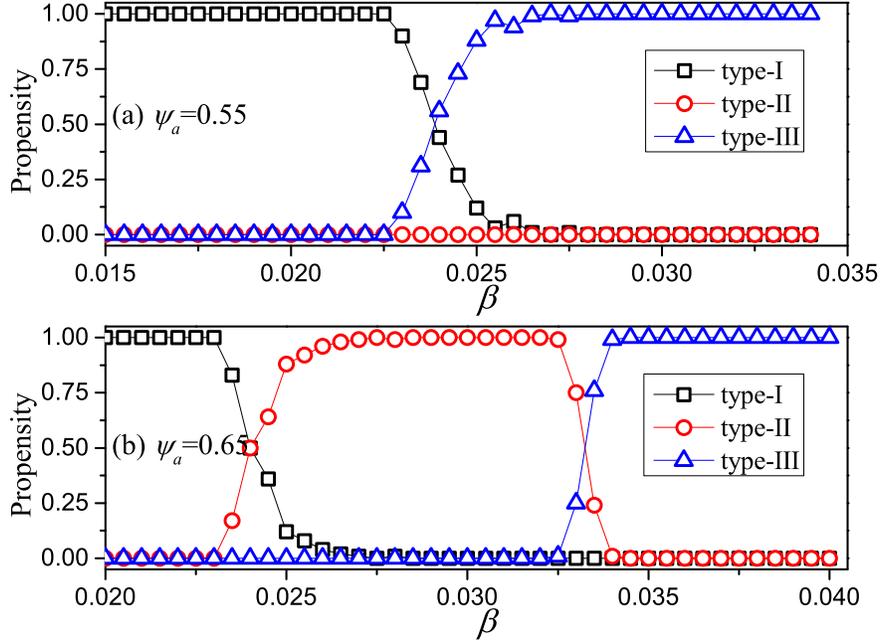,width=0.8\linewidth}
\caption{(Color online) Propensity of outcomes depending on infectivity ($\beta$) and heterogeneity ($\psi_{\rm a}$). Shown is the probability that a given type of behavior 
is observed when simulating a random initial network with the respective parameter values (see text). These probabilities where estimated by classifying the outcomes of 100 simulation runs for each parameter combination. For low values of heterogeneity (top, $\psi_a = 0.55$) we observe type I (recovery) behavior if infectivity is low and and type III (endemic) behavior if infectivity is high. At intermediate values there is a transition region where both outcomes are possible. For systems with strong heterogeneity (bottom, $\psi_a = 0.65$) additionaly type II (outbreak,collapse) behavior is observed at intermediate values of infectivity, which is separated from type I and type III behavior by two transition regions.    
 Parameters: $\psi_b = 0.05$, $w=0.2$, $ \mu=0.002 $, $i_0=0.0002$, $N = 10^5$, $K = 10^6$.}\label{Figure4}\end{center}
\end{figure}

We observe that in some ranges of infectivity different types of outcomes are possible. To explore this in more detail we use the proposed classification to plot the propensity of outcomes in Fig.~\ref{Figure4}. For low values of heterogeneity between nodes $\psi_a=0.55$ we find that for systems there are only two possible outcomes, namely type I (recovery) and type III (endemic) behavior. However, if the susceptibility of agents is very heterogeneous then also type II (outbreak,collapse) behavior is observed.

In Fig.~\ref{Figure4} we see that regions of different types of outcomes are separated by transition regions where 2 outcomes are possible. To prepare for the more detailed exploration below, let us now construct a 2-parameter phase diagram of the system (Fig.~\ref{Figure5}). In this diagram we draw the phase boundaries at the points where different type of outcome occurs in simulation, e.g.~the phase boundary between outcomes of type I and type II $\beta_{l}$ is set of points where type II outcome starts to show up and the same to phase boundary between type II and type III $\beta_{u}$.   

\section{Onset of Outbreaks \label{secPercolation}}
Let us now try to understand the phase diagram analytically. We start by considering the onset of outbreaks, i.e.~the boundary of type I behaviour. The ability of a disease to spread in a population can be quantified in terms of the basic reproductive number $R_0$, which denotes 
the number of secondary infections, caused by one infected, in the limit of low disease prevalence. If $R_0>1$ the disease can percolate through the network, and thus outbreaks become possible. 

\begin{figure}
\begin{center}
\epsfig{file=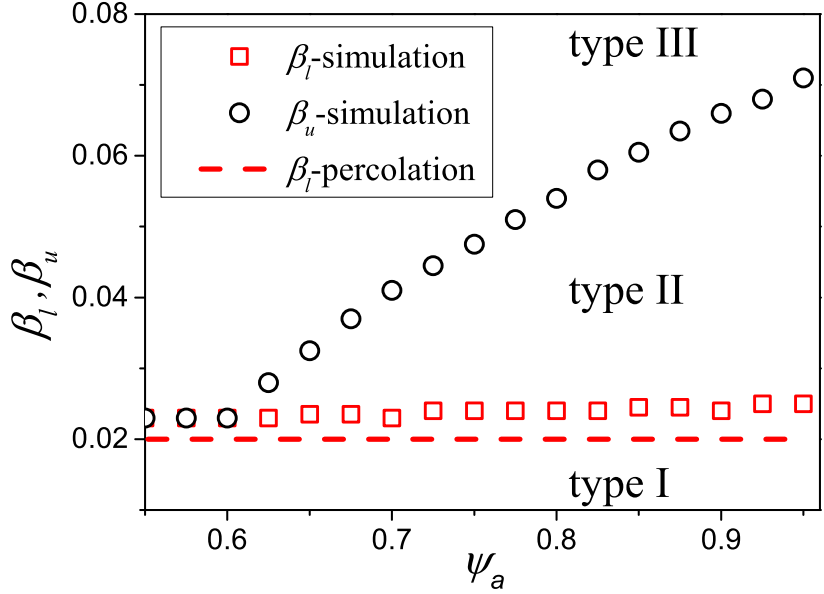,width=0.8\linewidth}
\caption{(Color online) Phase boundaries betweeen different types of outcomes; $\beta_l$ refers to the I/II boundary and $\beta_u$ to the II/III boundary. Shown are results from the classification of simulation runs (symbols) and an estimate using percolation theory from Eq.~\ref{eq_meanfield_threshold} (dashed line). Parameters: $\psi_b = 0.05$, $w=0.2$, $ \mu=0.002 $, $i_0=0.0002$, $N = 10^5$, $K = 10^6$.}\label{Figure5}
\end{center}
\end{figure} 

We can compute $R_0$ by considering a typical newly infected agent and computing the number 
of neighbours this agent will infect before recovering. 
Following~\cite{gross2006epidemic} we take into account that the number of links of the focal agent decreases in time as neighbouring agents rewire away. The loss rate of links is equal to the rewiring rate $\omega$. Thus the remaining degree after time $t$ is 
\begin{equation}
k(t)= k_0 {\rm e}^{-\omega t},
\end{equation}
where $k_0$ is the initial number of neighbours. Since we are interested in the limit of low prevalence, all neighbors can be assumed to be susceptible and we can find the number of secondary infections by multiplying the probability of transmission, which we call $p$ for the moment, and then integrating over the typical time to recovery $1/\mu$. This yields
\begin{equation}
\label{Rnaught}
R_0 = p \int_0^{1/\mu} k_0 {\rm e}^{-\omega t} {\rm dt} = \frac{pk_0}{\omega}\left(1-{\rm e}^{-\frac{\omega}{\mu}}\right).
\end{equation} 
For the heterogeneous network, we can express the probability of transmission $p$ as 
\begin{equation}
p=\beta (x_a \psi_a + x_b \psi_b)
\end{equation}
where $x_a$ is the probability that a randomly chosen neighbour is of type A, and $x_b$ the probability that a randomly chosen neighbour is of type B. As the initial network is an Erd\H{o}s-R\'{e}nyi random graph, $x_a = p_a$ and $x_b = p_b$. Substituting in to Eq.~(\ref{Rnaught}) and and setting $R_0=1$ yields
\begin{equation}
1=\frac{\beta (p_a\psi_a + p_b \psi_b )k}{\omega}\left(1-{\rm e}^{-\frac{\omega}{\mu}} \right)
\end{equation}    
and hence the threshold
\begin{equation}\label{eq_meanfield_threshold}
\beta_l=\frac{\omega}{k\langle \psi \rangle\left(1-{\rm e}^{-\omega/\mu} \right)},
\end{equation}
with $\langle \psi \rangle=p_a\psi_a+p_b\psi_b$. Expectedly this equation is very closely related to the epidemic threshold in the homogeneous system. The two values of $\psi$ are effectively averaged and only the numerical mean appears.

A comparison of the outbreak threshold identified based on percolation arguments and the numerical results show good qualitative agreement (Fig.~\ref{Figure5}). In the simulations we observe the outbreak only at slightly higher levels of infectivity, which is most likely a finite size effect. Closely above to the theoretical threshold for the infinite size system the finite size simulation can still collapse to the absorbing disease free state due to stochastic extinction.  

The results obtained above were based on the assumption that agents of type A and type B are well mixed. While this assumption is true in the initial state, very different outbreak thresholds can be found if the assumption is violated, for instance if rewiring in response to an earlier outbreak led to a non-random mixing in the population. We explore this particular scenario in detail in the next section. 

\begin{figure}
\begin{center}
\epsfig{file=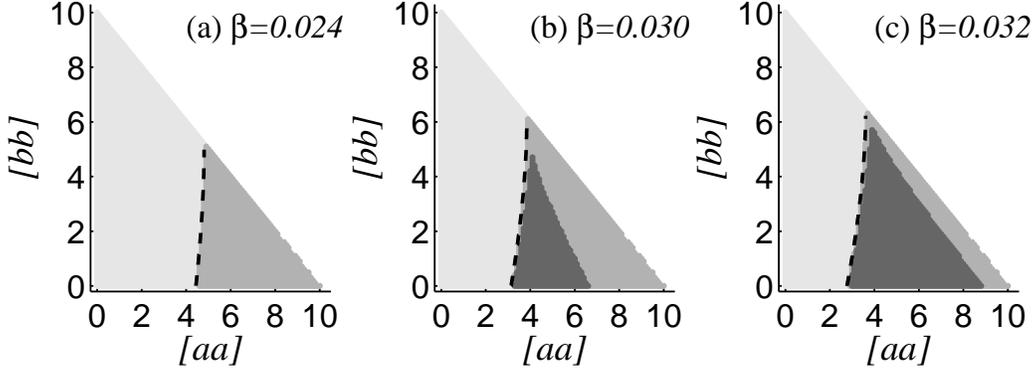,width=0.99\linewidth}
\caption{Impact of network structure in the initial state. Shown is the stability threshold found by percolation methods, Eq.~\ref{eq_threshold_link} (dashed line), in comparison to local asymptotic stability of the disease-free state computed based on the eigenvalues of the Jacobian matrix of the moment equations (Sec.~\ref{secMoments}). The figure shows the regions 
of stable disease free (type I, light grey), outbreak and collapse (type II, medium grey) and endemic (type III, dark grey) behavior, where we used simulations of the moment equations to distinguish between types II and III. In the remainder of the figure (white) no networks exist as the sum of a--a links and b--b links would be greater than the total number of links in the system. The figure shows that the agreement between the threshold for the onset of outbreaks computed by the two different approximations is almost perfect.  Parameters: $\psi_b = 0.05$, $\psi_a = 0.65$, $w=0.2$, $ \mu=0.002 $, $i_0=0.0002$, $N = 10^5$, $K = 10^6$ and $[aa]+[ab]+[bb]=\langle k \rangle /2$.}\label{Figure6}
\end{center}
\end{figure}

To gain a general understanding of the effects of assortativity in the disease free state let us now consider a disease free state with given number of a--a and b--b links. We denote the density of these links in the population by $[aa]$ and $[bb]$, respectively. The numerical values of both of these quantities are understood to be normalized with respect to the total number of nodes $N$. In this notation the density of a--b links $[ab]$ can then be computed from the conservation law
\begin{equation}
k = 2 ([aa]+[ab]+[bb])
\end{equation} 
Given $[aa]$ and $[bb]$ we can therefore write the number of nodes of types $i$ that are infected by a given node of type $j$ as 
\begin{equation}
R_{i,j}= \frac{\beta \psi_i [ij] (1+\delta_{i,j})}{\omega p_j}\left(1-{\rm e}^{-\frac{\omega}{\mu}}\right).
\end{equation}
These values form the entries in a $2\times 2$ next-generation matrix. The disease can spread if the leading eigenvalue of this matrix is larger than one. By pulling the repeated factor out of the matrix we get the condition
\begin{equation}\label{eq_threshold_link}
\lambda>\frac{\omega}{\beta\left(1-{\rm e}^{-\frac{\omega}{\mu}}\right)}
\end{equation} 
where $\lambda$ is the leading eigenvalue of 
\begin{equation}
{\rm \bf R}' = \left(\begin{array}{cc}
  \frac{2\psi_a [aa]}{p_a} & \frac{\psi_a [ab]}{p_b} \\
  \frac{\psi_b [ab]}{p_a} & \frac{2\psi_b [bb]}{p_b}
  \end{array} \right) 
\end{equation}
This provides a condition that can be solved for, say, the critical number of a--a links $[aa]$
at which outbreaks start. While easy to compute this condition is quite lengthy and is hence 
omitted here. The result is shown in  Fig.~{\ref{Figure6}}.

The computation shows that for a given value of infectivity an outbreak can occur if the density of a--a links is sufficiently high. This is intuitively reasonable as a disease close to the threshold will mainly spread in the highly-susceptible (type A) population.  

\section{Moment expansions \label{secMoments}}
To investigate the system further we can capture the dynamics by a moment expansion. Following the procedure in~\cite{gross2006epidemic,yang2015large} we write a system of differential equations that capture the dynamics of the abundances of different types of links and node states. We use symbols of the form $[X_u]$ and $[X_uY_v]$ with $X,Y\in\{{\rm I},{\rm S}\}$ and $u,v\in\{{\rm a},{\rm b}\}$ to respectively denote the proportion of agents and per capita density of links between agents of a given type. For instance $[I_{\rm a}]$ is the proportion of agents that are infected and of type A, and $[S_{\rm a}I_{\rm b}]$ is the per capita density of links between susceptible agents of type A and infected agents of type B. All of these variables are normalized with respect to the total number of nodes $N$. Given the number of infected nodes of a given type we can thus find the number of susceptible nodes by using the conservation law $[I_{u}]+[S_{u}]=p_{u}$. 

The time evolution of the proportion of nodes that are infected and of type A and B  can be respectively written as 
\begin{equation}
\frac{\rm d}{\rm dt}[I_{\rm a}]=-\mu [I_{\rm a}]+\beta \psi_{\rm a} \sum_v [S_{\rm a}I_v],
\end{equation}  
\begin{equation}
\frac{\rm d}{\rm dt}[I_{\rm b}]=-\mu [I_{\rm b}]+\beta \psi_{{\rm b}} \sum_v [S_{\rm b}I_v].
\end{equation}  

For the link densities, using a pair-approximation leads to equations of the form
\begin{equation}
\label{longEq}
\begin{split}
\frac{d[S_{\rm a}S_{\rm a}]}{dt}= \mu [S_{\rm a}I_{\rm a}] -2\beta\psi_{\rm a} (\frac{[S_{\rm a}S_{\rm a}][S_{\rm a}I_{\rm a}]}{[S_{\rm a}]}+\frac{[S_{\rm a}S_{\rm a}][S_{\rm a}I_{\rm b}]}{[S_{\rm a}]}) 
+\frac{\omega [S_{\rm a}]}{[S_{\rm a}]+[S_{\rm b}]} ([S_{\rm a}I_{\rm a}]+[S_{\rm a}I_{\rm b}]),
\end{split}
\end{equation}
where the terms on the right hand side describe the impact of the different processes on the motif considered, $[S_{\rm a}S_{\rm a}]$ in this example. For instance the first term corresponds to the creation of $S_{\rm a}$--$S_{\rm a}$ links 
due to recovery of the infected node in $S_{\rm a}$--$I_{\rm a}$ links. In total the $I_{\rm a}$ nodes recover at the rate $\mu [I_{\rm a}]$. Every such recovery event creates an expected number of $S_{\rm a}$--$S_{\rm a}$ links that is identical to the average number of $I_{\rm a}$--$S_{\rm a}$ links anchored on an $I_{\rm a}$ node, which is $[I_{\rm a}S_{\rm a}]/[I_{\rm a}]$. In summary, the change in the density of $S_{\rm a}$--$S_{\rm a}$ links due to recovery of $I_{\rm a}$ nodes is $\mu[I_{\rm a}][I_{\rm a}S_{\rm a}]/[I_{\rm a}]=\mu[I_{\rm a}S_{\rm a}]$, which explains the first term in Eq.~(\ref{longEq}). 

Similarly, 
\begin{equation}
\begin{split}
\frac{d[S_{\rm b}S_{\rm b}]}{dt}=& \mu [S_{\rm b}I_{\rm b}] -2\beta\psi_{\rm b} (\frac{[S_{\rm b}S_{\rm b}][S_{\rm b}I_{\rm a}]}{[S_{\rm b}]}+\frac{[S_{\rm b}S_{\rm b}][S_{\rm b}I_{\rm b}]}{[S_{\rm b}]})  \\
&+\frac{\omega [S_{\rm b}]}{[S_{\rm a}]+[S_{\rm b}]}([S_{\rm b}I_{\rm a}]+[S_{\rm b}I_{\rm b}]),
\end{split}
\end{equation}
\begin{equation}
\begin{split}
\frac{d[S_{\rm a}S_{\rm b}]}{dt}= & \mu ([S_{\rm b}I_{\rm a}]+[S_{\rm a}I_{\rm b}])  \\
&-\beta\psi_{\rm a} (\frac{[S_{\rm b}S_{\rm a}][S_{\rm a}I_{\rm a}]}{[S_{\rm a}]} 
+\frac{[S_{\rm b}S_{\rm a}][S_{\rm a}I_{\rm b}]}{[S_{\rm a}]}) 
-\beta\psi_{\rm b} (\frac{[S_{\rm a}S_{\rm b}][S_{\rm b}I_{\rm a}]}{[S_{\rm b}]}
+\frac{[S_{\rm a}S_{\rm b}][S_{\rm b}I_{\rm b}]}{[S_{\rm b}]})   \\
&+\frac{\omega [S_{\rm b}]}{[S_{\rm a}]+[S_{\rm b}]}([S_{\rm a}I_{\rm a}]+[S_{\rm a}I_{\rm b}]) 
+\frac{\omega [S_{\rm a}]}{[S_{\rm a}]+[S_{\rm b}]}([S_{\rm b}I_{\rm a}]+[S_{\rm b}I_{\rm b}]),
\end{split}
\end{equation}

\begin{equation}
\begin{split}
\frac{d[S_{\rm a}I_{\rm a}]}{dt}=&2\mu [I_{\rm a}I_{\rm a}]-(\mu +\beta \psi_{\rm a} + \omega)[S_{\rm a}I_{\rm a}] 
+2\beta\psi_{\rm a}(\frac{[S_{\rm a}S_{\rm a}][S_{\rm a}I_{\rm a}]}{[S_{\rm a}]}+\frac{[S_{\rm a}S_{\rm a}][S_{\rm a}I_{\rm b}]}{[S_{\rm a}]})  \\
&-\beta\psi_{\rm a}(\frac{[S_{\rm a}I_{\rm a}][S_{\rm a}I_{\rm a}]}{[S_{\rm a}]}
+\frac{[S_{\rm a}I_{\rm a}][S_{\rm a}I_{\rm b}]}{[S_{\rm a}]}),
\end{split}
\end{equation}
\begin{equation}
\begin{split}
\frac{d[S_{\rm b}I_{\rm b}]}{dt}= &2\mu [I_{\rm b}I_{\rm b}]-(\mu +\beta \psi_{\rm b} + \omega )[S_{\rm b}I_{\rm b}] 
+2\beta\psi_{\rm b}(\frac{[S_{\rm b}S_{\rm b}][S_{\rm b}I_{\rm a}]}{[S_{\rm b}]}+
\frac{[S_{\rm b}S_{\rm b}][S_{\rm b}I_{\rm b}]}{[S_{\rm b}]})  \\
&-\beta\psi_{\rm b}(\frac{[S_{\rm b}I_{\rm b}][S_{\rm b}I_{\rm a}]}{[S_{\rm b}]}
+\frac{[S_{\rm b}I_{\rm b}][S_{\rm b}I_{\rm b}]}{[S_{\rm b}]}),
\end{split}
\end{equation}
\begin{equation}
\begin{split}
\frac{d[S_{\rm a}I_{\rm b}]}{dt}= &\mu [I_{\rm a}I_{\rm b}]-(\mu +\beta \psi_{\rm a} + \omega)[S_{\rm a}I_{\rm b}]   
+\beta\psi_{\rm b}(\frac{[S_{\rm a}S_{\rm b}][S_{\rm b}I_{\rm a}]}{[S_{\rm b}]}+
\frac{[S_{\rm a}S_{\rm b}][S_{\rm b}I_{\rm b}]}{[S_{\rm b}]})  \\
&-\beta\psi_{\rm a}(\frac{[S_{\rm a}I_{\rm b}][S_{\rm a}I_{\rm a}]}{[S_{\rm a}]}
+\frac{[S_{\rm a}I_{\rm b}][S_{\rm a}I_{\rm b}]}{[S_{\rm a}]}),
\end{split}
\end{equation}
\begin{equation}
\begin{split}
\frac{d[S_{\rm b}I_{\rm a}]}{dt}= & \mu [I_{\rm a}I_{\rm b}]-(\mu +\beta \psi_{\rm b} + \omega)[S_{\rm b}I_{\rm a}]  
+\beta\psi_{\rm a}(\frac{[S_{\rm a}S_{\rm b}][S_{\rm a}I_{\rm a}]}{[S_{\rm a}]}+
\frac{[S_{\rm a}S_{\rm b}][S_{\rm a}I_{\rm b}]}{[S_{\rm a}]}) \\
&-\beta\psi_{\rm b}(\frac{[S_{\rm b}I_{\rm a}][S_{\rm b}I_{\rm a}]}{[S_{\rm b}]}
+\frac{[S_{\rm b}I_{\rm a}][S_{\rm b}I_{\rm b}]}{[S_{\rm b}]}),
\end{split}
\end{equation}

\begin{equation}
\begin{split}
\frac{d[I_{\rm a}I_{\rm a}]}{dt}= -2\mu [I_{\rm a}I_{\rm a}]+\beta \psi_{\rm a}[S_{\rm a}I_{\rm a}]   
+\beta\psi_{\rm a}(\frac{[S_{\rm a}I_{\rm a}][S_{\rm a}I_{\rm a}]}{[S_{\rm a}]}+
\frac{[S_{\rm a}I_{\rm a}][S_{\rm a}I_{\rm b}]}{[S_{\rm a}]}),
\end{split}
\end{equation}
\begin{equation}
\begin{split}
\frac{d[I_{\rm b}I_{\rm b}]}{dt}= -2\mu [I_{\rm b}I_{\rm b}]+\beta \psi_{\rm b}[S_{\rm b}I_{\rm b}] 
+\beta\psi_{\rm b}(\frac{[S_{\rm b}I_{\rm b}][S_{\rm b}I_{\rm a}]}{[S_{\rm b}]}+
\frac{[S_{\rm b}I_{\rm b}][S_{\rm b}I_{\rm b}]}{[S_{\rm b}]}).
\end{split}
\end{equation}   

\begin{figure}
\begin{center}
\epsfig{file=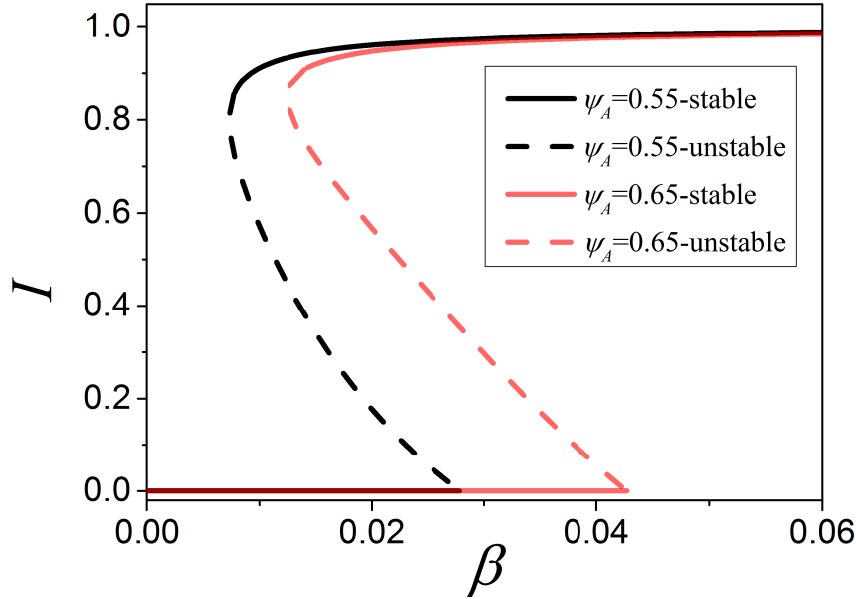,width=0.8\linewidth}
\caption{(Color online) Bifurcation diagram of the moment equations. Shown are branches of steady states for two values of heterogeneity, $\psi_A=0.55$ (black) and $\psi_A=0.65$ (red/gray). Numerical continuation reveals both stable (solid) and unstable (dashed branches).
Stability changes due to a transcritical (TC) and saddle-node bifurcations (SN). Between these two bifuractions a hysteresis loop is formed that is typical for adaptive SIS models.  
 Parameters: $ \psi_{\rm b} = 0.05 $, $\omega=0.2$, $ \mu=0.002 $, $ N=10^{5} $, $ K=10^{6} $.}\label{Figure7}
\end{center}
\end{figure}

In contrast to the percolation approach and agent-based simulations the moment expansion 
allows us to investigate the dynamics directly on an emergent level. In the context of the moment equations the different types of long-term behaviours now appear 
as attractors of a dynamical system. Numerical continuation reveals a bifurcation diagram that 
is typical of adaptive epidemic models (Fig.~\ref{Figure7}).

At sufficiently high infection rate, there is a stable steady state where the disease persists with high prevalence. When we gradually lower the infection rate this steady state becomes unstable due to a saddle node bifurcation, or by undergoing a Hopf bifurcation quickly followed by saddle-node bifurcation, depending on parameters. The limit cycle formed in the Hopf bifurcation only exists in a very small parameter range before it is destroyed in further bifurcations. 

The situation is more complex for the disease free states. While the branches of steady states where the disease is present have well-defined values in all of the dynamical variables, the disease free states form a manifold. All states in which the density of infected nodes is zero are necessarily stationary. However, this still permits networks with different values of the variables $[aa]$ and $[bb]$.

Above we already explored the stability of the manifold of disease-free steady states using the microscopic branching process approach. We can now replicate these results using the macroscopic moment expansion approach. For this purpose we compute the Jacobian matrix of the moment equations on the manifold of the disease-free steady states. These states are then stable if the leading eigenvalue of the Jacobian has a negative real part. Comparison of the threshold that is thus obtained with previous results (see Fig.~\ref{Figure6}) shows that the two approaches are in almost perfect agreement. 
   
\section{Transition to the endemic state \label{secTransition}}
Let us now turn our attention to the transition between type II and type III behavior. From the analysis above it is already evident that this transition is not caused by a local bifurcation. The endemic steady state is stable long before the destabilization of the disease-free state occurs. Therefore, the endemic state is an attractor throughout most of the parameter range considered here. For low values of heterogeneity, a system starting in the disease-free state approaches this attractor as soon as the disease-free state is destabilized. 

For higher values of heterogeneity the situation is different. The initial disease-free state is no longer in the basin of attraction of the endemic state. The system thus undergoes a single outbreak before it falls back to a different disease-free state (with a different distribution of links between node types) which is then stable against further outbreaks. 

The transition between type II (single outbreak) and type III (endemic state) behavior is represented by a transition of the initial disease-free saddle from one basin of attraction to another one. 
For a given parameter set we can visualize thee different basins of attraction based on numerical simulations (Fig.~\ref{Figure6}). We note that type II (outbreak-collapse) behavior occurs when the density of a--a links is high, whereas endemic behavior is observed for intermediate density of a--a links. While the a--a link density has to exceed a threshold value to allow outbreaks, the outbreak eventually collapses if a second threshold is exceeded. 

\begin{figure}
\begin{center}
\epsfig{file=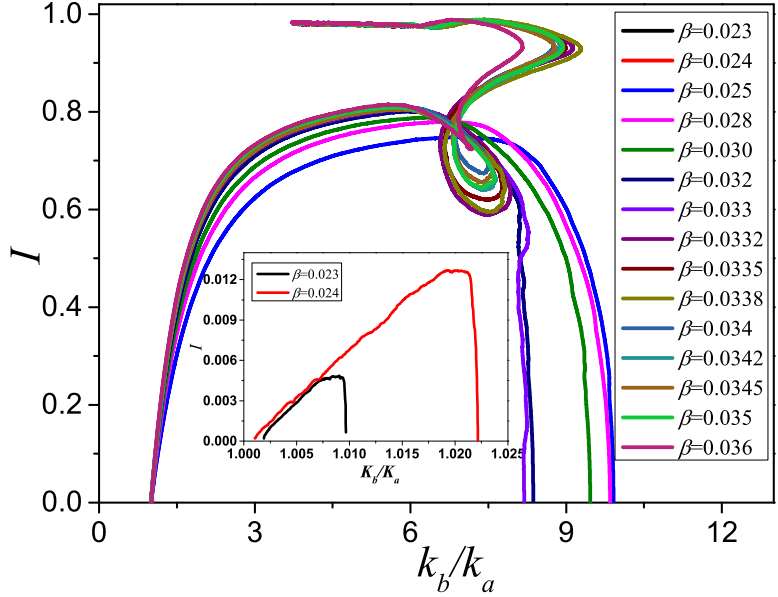,width=0.8\linewidth}
\caption{(Color online) Trajectories from agent-based simulation. Shown are 15 trajectories starting from the same initial state at different values of infectivity $\beta$. At low infectivity the trajectories remain in the vicinity of the initial state (inset). At higher infectivity there is an initial outbreak leading to high values of prevalence $I$ before collapsing back to a disease free state, where ratio between the degree of type B and type A nodes is now much higher than in the initial network. At even higher values of infectivity endemic behavior is observed as the system approaches a stable state with high prevalence. The transition to endemic behaviour occurs when trajectories encounter a point where the dynamics is almost stationary, which points to a heteroclinic bifurcation. Parameters: $\psi_a=0.65$, $\psi_b=0.05$, $w=0.2$, $ \mu=0.002 $, $i_0=0.0002$, $N = 10^5$, $K = 10^6$.}\label{Figure8}
\end{center}
\end{figure}
 
We note that outbreak (type II) dynamics always land the network in a final state that is characterized by lower connectivity of the highly-susceptible type A nodes, in which disease propagation is suppressed. Hence one can say that the outbreak inoculates the network against subsequent outbreaks of the same disease.  
 
The nature of the transition from type II to type III behavior is revealed when one considers 
trajectories from agent-based simulations (Fig.~\ref{Figure8}). As the parameter is tuned closer to the transition point the trajectories start to approach the saddle point that is formed in the fold bifurcation of the endemic state (see Fig.~\ref{Figure7}). In Fig.~\ref{Figure8} one can see one of the trajectories turning sharply in as it passes close to the saddle. This shows that the transition between type II and type III behaviour is caused by a saddle-heteroclinic bifuraction. In this bifurcation the unstable manifold from the saddle hits the initial state, such that a heteroclinic connection between saddles is formed. This connection also marks a basin boundary, such that in the bifurcation the initial state passes from one basin of attraction to the other. 

\begin{figure}
\begin{center}
\epsfig{file=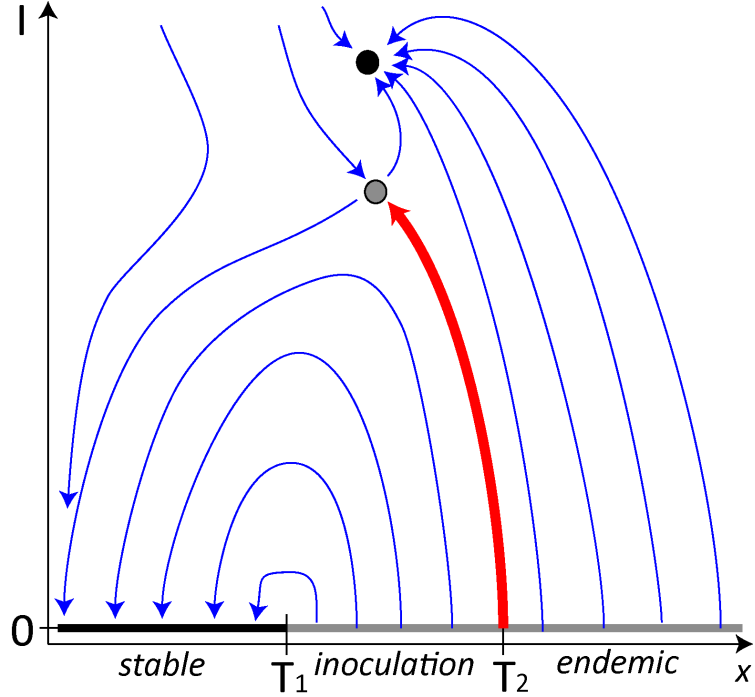,width=0.6\linewidth}
\caption{(Color Online) Simplified sketch of the phase portait in the epidemic model. Shown is a flow field (thin blue arrows) the attracting endemic state (black circle), a saddle point (grey circle) and a manifold of disease-free steady states (strong grey/black line), which can be stable (black) or unstable (grey). Depending on the initial value of the x-axis we can distinguish between stable disease-free (type I), outbreak and collapse (type II), and endemic (type III) behavior, indicated by labels on the axis. The behaviour changes at two threshold values ($T_1$, $T_2$) which are marked by a local change in the stability of the manifold and the heteroclinic connection.  We note that this sketch has been simplified from the situation in the epidemic model. If the x-axis were the a--a link density $[aa]$ the type II behavior would occur for intermidate values whereas the type III behavior would occur at high values, which is harder to visualize in a 2d-plot, but qualitatively simlar.\label{Figure9}}
\end{center}
\end{figure}

We can illustrate the situation with a simplified sketch of the phase portrait (Fig.~\ref{Figure9}). The figure shows how two thresholds divide the manifold of 
disease-free steady states into different sections in which perturbations lead to 
three different types of outcomes observed. If other parameters of the system change then these two thresholds move such that for a given initial condition, the transitions appear as transcritical and heteroclinic bifurcations respectively.   

Let us emphasize that the x-axis in Fig.~\ref{Figure9} cannot be the variable $[aa]$ as the different types of behaviour would occur in a different order (cf.~Fig.~\ref{Figure6}). The different order of sections when plotted over $[aa]$ does not imply qualitatively different dynamics, but is more difficult to visualize in a two-dimenional sketch.      
       
\section{Solvable stylized model \label{secSolvable}}    
Even the simplified ODE system discussed above has eleven degrees of freedom, and as such it is difficult to analyse in detail. In fact, the basic phenomenon of inoculation via a heteroclinic bifurcation can be captured in a solvable two-dimensional stylized model as we now describe. We consider a well-mixed population with two susceptible types (denoted $S_a$ and $S_b$ as previously), and a single infective type $I$. Having removing the network structure, to see the same phenomenon inoculation, it is necessary to introduce a new non-linear term to induce bistability. We keep the same infection as above, but make a minimal modification to recovery: instead of spontaneous recovery, infectious individuals may be coopted back to a susceptible state by interaction with a pair of susceptible individuals of the same type. 

While the cooption to the susceptible type may seem strange at first glance, very similar 
mechanisms are typically considered in threshold models of  opinion formation, including for instance an adaptive network model for opinion formation among locusts~\cite{huepe2011adaptive}. While we intend the proposed model mainly as an abstract illustration, one can imagine that very similar models can be relevant in situations where 
both opinion formation and epidemic processes occur. This is the case for instance, when choices can be made that prevent infection (e.g.~vaccination) or transmission (e.g.~hygiene, safer sex).   

The dynamics of the simplified model are captured by the rate equations
\begin{equation}
\begin{split}
\frac{d[S_a]}{dt}=&-\beta \psi_a[I][S_a]+\mu[I][S_a]^2\\
\frac{d[S_b]}{dt}=&-\beta \psi_b[I][S_b]+\mu[I][S_b]^2\\
\frac{d[I]}{dt}=&\beta [I]\Big(\psi_a[S_a]+\beta \psi_b[S_b]\Big)-\mu[I]\Big([S_a]^2+[S_b]^2\Big)\,.
\end{split}
\end{equation}

\begin{figure}
\begin{center}
\epsfig{file=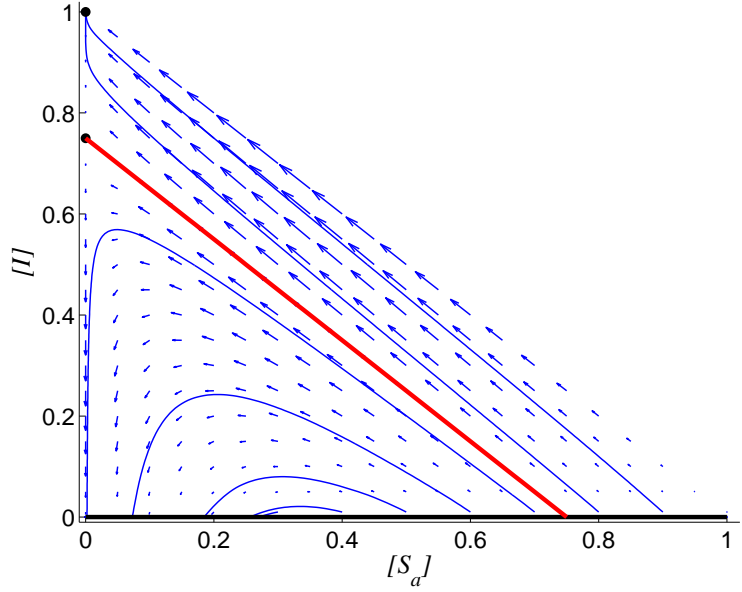,width=0.6\linewidth}
\caption{(Color online) Phase portrait for the simplified model. The phase portrait contains a manifold of steady states (strong black line) at zero prevalence. In addition there are two steady states at non-zero prevalence (black dots). The lower of these two states is a saddle whose unstable manifold (red line) forms the separatrix between outbreak and endemic behavior. This is illustrated by the flow field (blue arrows) and example trajectories (thin blue lines).  Parameters: $\beta=0.5$, $\mu=0.5$, $\psi_a=1$, $\psi_b=0.25$.
\label{Figure10}} \end{center}
\end{figure}

Note that the system is two-dimensional since $[S_a]+[S_b]+[I]=1$ is a conserved quantity. The line $[I]=0$ is a manifold of fixed points. Along the absorbing lines $[S_a]=0$ and $[S_b]=0$, the system is reduced to the one-dimensional ODE
\begin{equation}
\frac{d[I]}{dt}=\beta \psi_{\ast}[I](1-[I])-\mu[I](1-[I])^2\,,
\end{equation}
where $\ast\in\{a,b\}$. The behaviour of this system has two phases. There are always steady states at $[I]=0$ (extinction) and $[I]=1$ (endemic infection), with the possibility of a third at $[I]=1-\beta\psi_\ast/\mu$. If this third steady state lies in $(0,1)$ then it is a saddle, and the extinct and endemic states are stable. If it lies outside the physically relevant region then the extinct state is unstable. 

By choosing $\psi_b<\psi_a$ appropriately, we are able to realise a situation in which there is a saddle on the $[S_a]=0$ line but not on $[S_b]=0$. This structure motivates the unusual non-linear choice made for recovery.

The phase portrait of the system is shown in Fig.~\ref{Figure10}. From the figure, perturbation around a state with $[I] = 0$ has three possible outcomes. For small $[S_a]$, we have a type I region, where no outbreaks can occur. For large $[S_a]$, the trajectory is carried all the way to the stable endemic equilibrium at $[I]=1$ in a type III scenario. In between, there is a range of values for $[S_a]$ with type II trajectories that initially depart, but then return to the $[I]=0$ line. This region is bounded on the left by the point where the non-zero eigenvalue of the Jacobian matrix changes sign, which we compute to be the point where $[S_a]$ solves
\begin{equation}
0=\beta\psi_a[S_a]+\beta\psi_b(1-[S_a])-\mu[S_a]^2-\mu(1-[S_a])^2\,.
\end{equation}
On the right the type II region is bounded by the separatrix of the endemic and extinct states, which can be found by examining  
\begin{equation}
\frac{d[I]}{d[S_a]}=-1+\frac{1-[I]-[S_a]}{(-\beta\psi_a+\mu [S_a])[S_a]}\Big(-\mu(1-[I]-[S_a])+\beta\psi_b\Big)\,,
\end{equation}
implying the separatrix $[I]=1-[S_a]-\beta\psi_b/\mu$.

\begin{figure}
\begin{center}
\epsfig{file=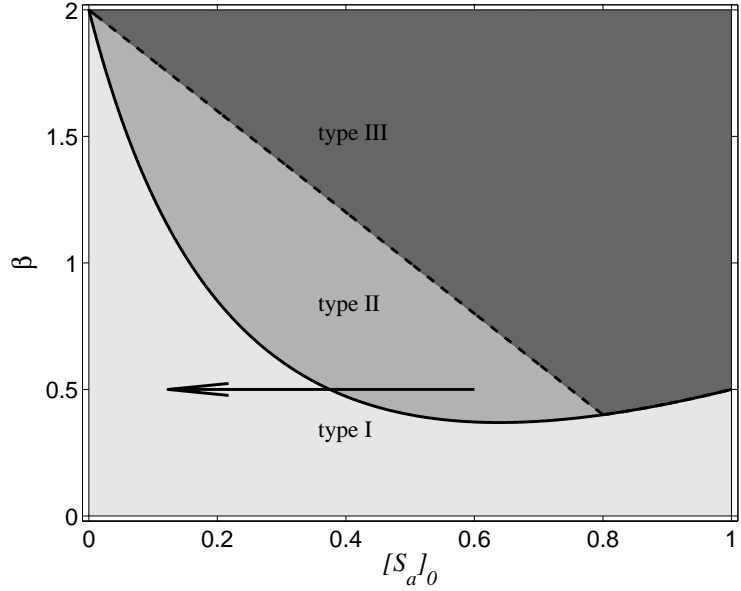,width=0.6\linewidth}
\caption{Phase diagram of the simplified model. Transcritical (solid line) and heteroclinic (dashed line) bifurcations separated phases of qualitatively different behavior: type I (disease free, light grey), type II (outbreak and collapse, medium grey), type III (endemic, dark grey). Outbreaks take the system from the type II region into the type I region (black arrow) and thus inoculate it against further outbreaks. Parameters: $\mu=0.5$, $\psi_a=1$, $\psi_b=0.25$. 
}\label{Figure11}\end{center}
\end{figure}

The results above allow us also to draw a phase diagram of the system (Fig.~\ref{Figure11}). In this diagram stable disease-free behavior (type I) is separated from epidemic behvior (type II and III) by a transcritical bifurcation, while outbreak (type II) and epidemic (type III) behaviour are separated by the heteroclinic bifucation. 

Trajectories starting in the type II phase lead to final states in the type I phase. In fact, 
the black arrow is the trajectory for $\beta = 0.5$, $[S_a]_{0} = 0.6$. Again, we can think of this kind of event as an inoculation, since the initial outbreak is crushed, and we are left with fewer type A susceptibles so that future outbreaks need a much higher $\beta$ (around 1.8 in this case) to succeed.
 
\section{Conclusions}
In this paper we investigated a previously proposed model for the spreading of a disease across a network in the face of behavioral responses to the disease and intra-individual heterogeneity of epidemic parameters. To understand the dynamics of this system we used a variety of tools, including agent-based simulation, percolation theory, moment expansions, 
analytical bifurcation theory, numerical integration of ODEs and continuation. 

Our results point to a phenomenon that we named \emph{network inoculation}. Introducing a disease into a given network may lead to an outbreak that collapses and leaves the network 
with a different topology as agents have rewired their connections in response to the disease. Although the altered topology will be generally more heterogeneous than the initial topology, it is more resilient to disease outbreaks. In this sense network inoculation is strongly reminiscent of immunological inoculation as in both cases contact to the pathogen leads to a response that hardens the system against future exposure to the pathogen.

Our analysis showed that the outbreak and collapse dynamics characteristic of network inoculation occurs in a region bordered by two phase transitions. When viewed from a macroscopic perspective one of these transitions is a transcritical bifurcation, whereas the 
other is a saddle-heteroclinic bifurcation. Network inoculation thus provides a (rare) example of a phenomenon where a global bifurcation causes a phase transition in a model that can be understood both on the micro- and macroscale.

We emphasize that network inoculation is not a peculiarity of the specific model studied here. By contrast, we expect the phenomenon to occur in a wide variety of models as soon as certain requirements are met. While the phenomenon may as well occur in other models, let us for consistency summarise the requirements of network inoculation in epidemic terms. Network inoculation can occur if there is 
\begin{enumerate}
\item{} A disease-free attractor (inoculated outcome)
\item{} An endemic attractor (endemic outcome)
\item{} A variety of unstable disease-free states (initial states)   
\end{enumerate}    
The actual inocculation strictly-speaking only requires condition 1 and 3, whereas condition 2 makes the onset of inoculation via a heteroclinic bifurcation possible.  
 
If the first two conditions are met there will be generally a saddle of some sort whose stable manifold marks the separatrix between the basins of the two attractors. Network inoculation will occur if the initial state is in (or on) the basin of the inoculated outcome. When parameters are changed the separatrix will generally move, which can cause an initial state to enter or leave the basin of the inoculated outcome, in a heteroclinic bifurcation. 

The conditions above require a bistability between an endemic (1) and a disease-free (2) state. While such bistability is not observed in the most simple models, it is very common 
in even slightly more complex models. In particular this bistability has been observed in numerous variants of the adaptive SIS models. It therefore seems to be a robust feature of epidemiological models that appears once behavioral responses to the disease are modeled. 

Furthermore we require the existence of multiple disease-free states with different stability properties. While the simplest epidemiological models have only a single disease-free state multiple disease free states naturally appear as soon as an additional macroscopic variable exists. 

Network inoculation was not observed in previous investigations of the adaptive SIS models. While this model shows robust bistability it has only a unique disease free state and hence does not meet the requirements of network inoculation. Likewise, network inoculation was not observed in previous models of epidemics in heterogeneous populations. In these models there are naturally multiple disease-free states which differ in the connectivity of the different classes of individuals. However, because these previous models did not consider adaptive rewiring of links the connectivities of the different classes of agents are parameters, rather than dynamical variables. Thus the different disease-free states are not observed simultaneously for one choice of parameters, hence again inoculation-type dynamics cannot occur. 

Once intra-individual heterogeneity and adaptive network rewiring are both considered multiple disease free states that differ in the connectivity of classes of individuals occur robustly. 
Because adaptive rewiring can change these connectivities, they are now dynamical variables, and the multiple disease-free states can be observed simultaneously, for a given set of the remaining parameters. When multiple disease-free states exist the generic expectation is that they will have different stability properties at least in some reason of the parameter space, and thus there will in general be a parameter region where the conditions for network inoculation in the narrow sense are met. 

Because bistability between endemic and disease free states has proven to be a very robust feature of adaptive epidemiological models, we can moreover expect the onset of network inoculation via the heteroclinic bifurcation to be a common phenomenon. Both ingredients, the adaptive response of the network to the disease, and intra-individual heterogeneity are known to exist in the real world. In the light of the arguments above we expect network inoculation, and its onset via the heteroclinic bifurcation to occur whenever these two ingredients are combined in the same model. 

Thus it seems that the reason why network inoculation has not been observed in the past is not the phenomenon itself is rare, but rather that the models that have been studied so far have been too strongly simplified to capture this, potentially common, phenomenon.

\acknowledgements

This work was supported by the EPSRC under grant EP/K031686/1, National Natural Science Foundation of China (Grant Nos. 11575041, 61433014), and the Program of Outstanding Ph.D. Candidate in Academic Research by UESTC (Grant No. YBXSZC20131036). TR gratefully acknowledges the support of the Royal Society.

\end{document}